\documentstyle[12pt]{article}

\let\a=\alpha	\let\b=\beta	\let\g=\gamma
\let\d=\delta	\let\e=\epsilon 
	\let\q=\theta	
 \let\l=\lambda	\let\m=\mu
\let\n=\nu	\let\x=\xi		\let\r=\rho
\let\s=\sigma		
\let\f=\phi	\let\c=\chi	
	\let\G=\Gamma	\let\D=\Delta

\let\F=\Phi	 
 \let\la=\label	\let\ci=\cite

\def\bd{\begin{document}}
\def\ed{\end{document}}
\def\ds{\documentstyle}	\let\fr=\frac
\let\bl=\bigl	\let\br=\bigr \let\Br=\Bigr
\let\Bl=\Bigl \let\bm=\bibitem
\let\na=\nabla \let\pa=\partial
\let\ov=\overline
\newcommand{\be}{\begin{equation}}
\newcommand{\ee}{\end{equation}}
\def\ba{\begin{array}} \def\ea{\end{array}}
\newcommand{\ho}[1]{$\,	^{#1}$}
\newcommand{\hoch}[1]{$\,	^{#1}$}
\newcommand{\bea}{\begin{eqnarray}}
\newcommand{\eea}{\end{eqnarray}}
\newcommand{\ra}{\rightarrow}
\newcommand{\lra}{\longrightarrow}
\newcommand{\Lra}{\Leftrightarrow}
\newcommand{\ap}{\alpha^\prime}
\newcommand{\bp}{\beta^\prime}
\newcommand{\tr}{{\rm	tr}	}
\newcommand{\Tr}{{\rm	Tr}	}
\newcommand{\NP}{Nucl.	Phys.	}
\newcommand{\tamphys}{\it	Center	for
Theoretical	Physics\\ Physics	Department	\\
Texas	A	\&	M	University \\	College	Station,
Texas	77843}

\begin{document}

\hfill{CTP-TAMU-47/93}

\hfill{hep-ph/9311305}

\vspace{24pt}

\begin{center}
 { \Large  {\bf  Supersymmetry
Anomalies, the Witten Index and
 the Standard Model} }

\vspace{12pt}

{\bf John Dixon}\footnote{Supported in part by the U.S. Dept of
Energy, under grant DE-FG05-91ER40633, Email:
dixon@phys.tamu.edu}

\vspace{12pt}

\tamphys

\vspace{6pt}

 \vspace{6pt}

{\bf Abstract}

\end{center}

{\small The supersymmetric standard model (SSM) contains a
wealth of potential supersymmetry anomalies, all of which occur in
the renormalization of composite operators of the theory. The
coefficients of the weak-E.M.  superanomalies  should be related
to the   Witten indices  of the  neutrino and photon
superfields, and the coefficients of the strong  superanomalies
should be related to the   Witten indices of the gluon and
photon superfields.  Assuming   the coefficients are non-zero, the
superanomalies  break supersymmetry in observable states.
However the neutral Higgs particles should remain in a
supermultiplet  because the Higgs supermultiplet  is not coupled to
any massless superfield in the SSM.  Assuming that  the overall
Witten index is non-zero,  supersymmetry is broken by
superanomalies and yet the vacuum remains supersymmetric.
This means that   the cosmological constant is naturally zero   after
supersymmetry breaking, even beyond perturbation theory. }

\section{Introduction}

All $N=1$ supersymmetric theories in four dimensions that have
chiral matter  have   potential
supersymmetry anomalies that can arise in the renormalization of
certain composite operators constructed in the theories.
However the action itself, at least in non-exotic theories, cannot
have any supersymmetry anomalies.  These statements are
consequences of the structure of the BRS cohomology of
supersymmetry which has been explored in references \ci{pss}
\ci{bdk} \ci{b} \ci{d}
\ci{dm} \ci{dmr}.  This situation is more or less the opposite of what
happens for the  gauge   and gravititational anomalies.

The basic principle obtained from the cohomology is very
simple. The simplest   composite operators that can develop
superanomalies are the composite antichiral spinor superfields and
the corresponding
 anomalies are   products of the elementary chiral
superfields in the theory, without any derivatives.  The details
of this structure depend on the details of the representations and
the gauge structure of the theory.    An
examination of a number of examples in simple models indicates that
the coefficients of such superanomalies are all  very likely to be zero,
except possibly in the case where the  theory has  non-Abelian gauge
fields as well as chiral superfields,  spontaneous breaking of the gauge
symmetry  (but not necessarily  the supersymmetry),  complex
representations of the gauge group  preventing  bare
mass terms for the chiral matter, and plenty of massless superfields
after gauge symmetry breaking \ci{tamu46}. The standard
supersymmetric model has all these properties, so one is led to
look at it for superanomalies, after trying simpler possibilities
without success.

The ideal would be to simply   calculate the
coefficients for some examples to answer this
question.  But this is hard.  It seems to be
necessary to first pick a model with a lot of
structure, then find some examples of
potentially  superanomalous operators, then
test the BRS identities in the relevant sectors, and finally remove
any superanomalies that  vanish by the   field
equations.   All possible  examples involve a lot of operator mixing,
and a calculation needs to be guided by  some
theoretical ideas about how the anomalies could
arise.  It now seems clear that the Witten index
and consequently zero mass superfields should
play an essential role, as we discuss in section
\ref{index} below.

As stated above, the composite operators that are susceptible to
superanomalies are the composite antichiral spinor superfields,
which we will denote by $\F_{\a}$.  The antichiral constraint is
$  {\cal D}_{\a} \F_{\b} = 0 $ where ${\cal D}_{\a} = \fr{\pa}{\pa
\q^{\a} } +
\fr{1}{2} \s^{\m}_{\a \dot \b} {\ov \q}^{\dot \b} \pa_{\m}
$ is the superspace chiral derivitive.  We
  can define components by the expansion $ {\hat
\F}_{\a}(x, \q, {\ov \q})  =  \f({\ov y})_{\a  }  + W({\ov y})_{\a \dot
\b} {\ov \q}^{\dot \b}
 + \fr{1}{2} {\ov \q}^2 \c({\ov y})_{\a  }   $
where the chirally translated spacetime variable
$y^{\m} = x^{\m} + \fr{1}{2}   \q^{\a} \s^{\m}_{\a \dot \b} {\ov
\q}^{\dot \b}
$
satisfies the equation
$ {\cal D}_{\a}  {\ov y}^{\m} = 0
$
Of course each of these components $\f, W, \x$ is a composite
field made
 from
the  component fields of the elementary superfields in the theory.

 The BRS cohomology tells us that these operators are subject
to     superanomalies  of the form $ m^q {\ov S} c_{\a} $ where
$c_{\a}$ is the commuting spacetime independent supersymmetry
ghost, and  the composite antichiral scalar superfield
${\ov S}$ (satisfying the constraint
 ${  D}_{\a} {\ov S} = 0 $) consists of
 a  product of the elementary antichiral superfields of the theory,
with no derivatives or superderivatives in it.  The mass
parameter of the theory is $m$, and $q$ is a power determined by
matching
 the dimensions of $\F_{\a}$ and ${\ov S} c_{\a}$ for each case.
Evidently  $\F_{\a}$ and $m^q {\ov S} c_{\a}$ must have identical
values for any   conserved    quantities
such as lepton number, charge, baryon number, mass dimension,
hypercharge, isospin and  colour charge.  Hypercharge and weak
isospin are spontaneously broken down to $U(1)_{EM}$ of course
in the SSM,  which must be properly taken into
account.

To calculate the superanomalies, one would
couple these operators to  a non-composite antichiral spinor
superfield source $\F'_{\a}$ with components
${\hat  \F'}_{\a} =  \f'_{\a}  + W'_{\a \dot \b} {\ov \q}^{\dot \b}
 + {\ov \q}^2 \c'_{\a  }  $  in the form ${\rm Action}_{\F} = \int
d^4 x d^2 {\ov \q} \F'^{\a} \F_{\a}
$.  Then   the anomaly would appear in the form
\be  \d \G_{\F} = e_1 \int d^4 x d^2 {\ov \q} \F'^{\a} c_{\a} m^q {\ov
S}
\ee
where $\G$ is the one-loop 1PI generating functional and
  $e_1$, the coefficient of the anomaly, is calculable in one-loop
perturbation theory.

Let us treat the composite superfields $\F_{\a}$ and $e_1  m^q
{\ov S}$ for a moment as if they were elementary fields with
canonical mass dimensions. Then we could eliminate the
supersymmetry anomalies by defining a new local nilpotent
transformation on these elementary fields as follows:
\be
\d' \F_{\a} =  {\ov S} c_{\a} +( c^{\b} Q_{\b} +{\ov  c}^{\dot \b}
{\ov Q}_{\dot \b}) \F_{\a}
\la{1}
\ee
\be
\d' \F'_{\a} =   ( c^{\b} Q_{\b} +{\ov  c}^{\dot \b}  {\ov Q}_{\dot \b})
\F'_{\a}
\la{2}
\ee \be
\d' { S}  =  ( c^{\b} Q_{\b} +{\ov  c}^{\dot
\b}  {\ov Q}_{\dot \b}) S
\la{3}
\ee

With this transformation the anomaly ceases to
be an anomaly
because it is now the variation of a local term:
\be    \int d^4 x d^2 {\ov \q} \F'^{\a} c_{\a}
{\ov S}
= \d'  \int d^4 x d^2 {\ov \q} \F'^{\a} \F_{\a}
\ee
However now one has to deal with the new
term in the `anomalous' super-BRS algebra in (\ref{1}).
The zero ghost charge invariants of the
transformations (\ref{1}),(\ref{2}),(\ref{3})
include an explicit dependence on $\q$. For
example we have:
\be
\d' {\rm Action}_{\rm anom} = \d' \int d^4 x d^4 \q  \F'_{\a} {\ov
{\cal  D}}^2  [\F^{\a} + \q^{\a} {\ov S}] =0
\la{anomaction}
\ee
Presumably,  this supersymmetry violating
$\q$-dependence means that the anomalies induce
 an effective action for the composite fields that
violates supersymmetry--the term (\ref{anomaction})
 is supersymmetry violating
because
$\d  {\rm Action}_{\rm anom} \neq 0$ where $\d$ is the usual
supersymmetry BRS operator.  It may be possible to restore the
supersymmetry by  some   non-local field redefinition, but then one
is led back to the supersymmetry anomaly.  Surely, this is what
one should expect--the whole point of doing local
BRS cohomology is to find out what cannot be
eliminated by local field redefinitions and local
renormalizations. An anomaly violates the
supersymmetry in a way that is not removable and
which should be physically significant.  This leads to the
conjecture that supersymmetry breaks itself in observable states
through calculable supersymmetry anomalies that arise at one loop
in  perturbation theory.

An  interesting problem arises here--to fully  analyze
the quadratic form of this action to see what the
mass eigenstates are and to see whether they do
break supersymmetry in an interesting way.   This
question is important even if all the
coefficients of the superanomalies are zero to
all orders of perturbation theory, because
anomalies are also of interest in the
non-perturbative regime--as is well known in
instanton physics.

Sometimes we will allow $ {\hat \F'}_{\a}$ to have weak isospin or
U(1) hypercharge.  For example for the case of example L2 of
Table \ref{examples}, we get the expansion:
 \be \d \G_{\F}  = c_1 m^4
 \int d^4 x    c_{\a}  [  \c'^{i \a}     {\ov L}_{  i}  + W'^{i \a \dot \b}
{\ov l}_{  i
\dot \b}   + \f'^{i \a}     {\check {\ov L}}_{  i} ]
 \la{lepanom1}  \ee

Allowing these indices on $\F'_{\a}$ would probably be wrong if we
wanted to treat the source $\F'_{\a}$ as a fundamental field, but
$\F'_{\a}$ in the present context is  {\em not}  a fundamental field.

\section{Supersymmetric Standard Model}
The chiral superfields of the supersymmetric
standard model (SSM) are summarized    in
Table \ref{ssm} .   But we  will not add any explicit
supersymmetry breaking terms,  the hope being that such terms
are not actually needed because supersymmetry breaks itself in
physical states in a way that does not show up in the
fundamental action.  It is worth noting   that if explicit
(`soft' or otherwise) supersymmetry breaking terms {\em are} put
into the action, they are very likely to    create interesting
problems because of the   superanomalies discussed here.

The   electromagnetic charge
 is $ {\cal Q} = {\cal I}_3 +
\fr{{\cal Y}}{2}  $ The  components of a    chiral superfield
are denoted
$ {\hat J}^i =
 J^i + \q^{\a} j_{\a} + \fr{\q^2}{2}
 {\check J}^i $.  Here ${\hat J}^i$ is the
chiral superfield,  $J^i$ is the scalar,
$ j_{\a}$ the spinor and ${\check J}^i$ the
auxiliary component.  Weak isospin is ${\cal I}$, weak hypercharge
is
${\cal Y}$,  baryon number is         ${\cal B}$ and lepton number is $
{\cal L}$.

\begin{table}
$\begin{array}{cccccccccccc}    {\rm Field}  &  SU(3) &
{\cal I}  & {\cal Y} & {\cal B} & {\cal L}
 & {\rm Field}  &  SU(3) &
{\cal I}  & {\cal Y}  & {\cal B}
 & {\cal L}\\
  {\hat J}^i     &
0         &  \fr{1}{2} &  -1 & 0 & 0
& {\hat {\ov J}}_i     & 0         &
\fr{1}{2} &   1 & 0 & 0
 \\ {\hat
K}^i  & 0 & \fr{1}{2} &  +1 & 0 & 0
& {\hat {\ov K}}_i  & 0 &
\fr{1}{2} &  -1 & 0 & 0  \\  {\hat M}
&  0 & 0 & 0 & 0 & 0 &
{\hat {\ov M}}  &  0 &
0 & 0 & 0 & 0 \\ {\hat L}^i& 0 &
\fr{1}{2} &  - 1& 0 & 1 & {\hat {\ov L}}_{  i}& 0 &
\fr{1}{2} &    1& 0 & - 1 \\
{\hat E}_R & 0 &
\fr{1}{2} &  +2 & 0 & -1 &
{\hat {\ov E}}_R & 0
& \fr{1}{2} &  -2 & 0 &   1 \\
 {\hat Q}^{i c} & 3 &
\fr{1}{2} &  \fr{1}{3}& \fr{1}{3} & 0
& {\hat {\ov Q}}_{  i
c}  & {\ov 3} & \fr{1}{2} &  \fr{-1}{3}&
\fr{-1}{3} & 0  \\
 {\hat
U}_{R c} & {\ov 3} & 0 & \fr{-4}{3} & \fr{-1}{3}
& 0 &{\hat {\ov U}}_{R}^{c} & {  3} & 0 &
\fr{ 4}{3} & \fr{1}{3}  & 0 \\
   {\hat D}_{R c}  & {\ov 3} & 0 &  \fr{2}{3}
& \fr{-1}{3}  & 0  &
{\hat {\ov D}}_{R }^{c}
& { 3} & 0 &  \fr{-2}{3} & \fr{1}{3}  & 0
\end{array}$
\caption{ Chiral superfields of the SSM}
\la{ssm}
\end{table}

In Table \ref{ssm},  $c=1,2,3$ is an index labelling the $3$ of color
SU(3),
$i=1,2$ is an isospin $\fr{1}{2}$ index, $\a=1,2$ is a two component
Weyl spinor index,
${\hat Q}^{1 a} = {\hat U}_L^{   a} $ and ${\hat Q}^{2 a} = {\hat
D}_L^{   a}
$ are  the left-handed  chiral superfields that contain   the up
and down quarks,   and  ${\hat {\ov Q}}_{ i  a }$, which transforms
as a
$\ov 3$ of SU(3), is the complex
conjugate  of ${\hat { Q}}^{ i  a }$.
 The  Yukawa
interactions are: \be  \int d^4 x d^2 \q \{  g_1 {\hat J}^i {\hat L}_i
{\hat E}_R
 + g_2 {\hat J}^i {\hat K}_i {\hat M}   + g_3
m^2 {\hat M}
 + g_4 {\hat Q}^{i a} {\hat K}_i {\hat U}_{R  a} + g_5 {\hat Q}^{i a}
{\hat J}_i {\hat  D}_{R  a}
\ee
 Our notation for the (gluon) vector superfields, in
the Wess-Zumino gauge,  is ${\hat G}^{a}_{\; \; b}   = [ \q^{\a}
\s^{\m}_{\a \dot \b} {\ov \q}^{\dot \b}   G_{\m}
+ \q^2   {\ov \q}^{\dot \a} {\ov g}_{\dot \a}
+    {\ov \q}^2 \q^{\a} g_{  \a}  +
\fr{1}{2} {\ov \q}^2 \q^2 {\check G} ]^{a}_{\; \; b}
  $.
and the gauge-covariant   chiral spinor
superfield is $
{\hat G}^{a  }_{\a\; \; b}    = [ {\ov {\cal D}}^2  (e^{-
{\hat G} } {\cal D}_{\a} e^{{\hat G} } )  ]^{a}_{\; \; b}$,
 which has the
expansion  ${\hat G}^{a  }_{\a\; \; b} = [ g_{\a}  +
\s^{\m \n}_{\a \b} \q^{\b}
  G_{\m \n}
 + {\check G} \q_{\a} +  \s^{\m}_{\a \dot \b}
D_{\m}  {\ov g}^{\dot \b  } ]^{a}_{\; \; b}  $.
 The $SU(2)$ ${\hat W}^i_{\;\;j}$ and $U(1)$ ${\hat V}$  gauge fields
are defined with a similar notation.
  Usually   the  $e^V$, $  e^G$ and $
e^W$ factors will be omitted with the
understanding that they are to be supplied as
necessary to make the gauge invariance work
properly.

As is discussed in \ci{haberkane}, there are
three `Higgs' fields $J, K$ and $M$, so that the
gauge symmetry naturally breaks from
$SU(2)  \times U(1)$ down to
$U(1)_{\rm E.M.}$ This set of Higgs fields also gives both the up and
down quark superfields a mass, which is not possible in the SSM
with just one weak isospin doublet Higgs multiplet.
The electrically neutral superfields  ${\hat J}$ and ${\hat K}$ develop
equal VEVs $ m  h = < J^1> = <K^2> $
where
$
h =  \sqrt{\fr{ g_3}{g_2} }
$.
This gauge breaking leaves the vacuum energy zero
of  course, because supersymmetry is not
spontaneously  broken
in this theory--all auxiliary fields have zero VEV.
\section{Equations of Motion}
One more point must be made before we turn to the
superanomalous operators.  An `anomaly' is not an anomaly if it
vanishes
when the equations of motion are used.   More
accurately, one needs to know the BRS cohomology
of the full BRS operator which includes the
variations of the antifields, which include the
 equations of motion of the fields.     The
antifields are introduced into the action with
terms like $ \int d^4 x d^2 \q  {\hat {\tilde L}}_i \d {\hat { L}}^i
$, where ${\hat {\tilde L}}_i$ is the anti-super-field that couples to
the left-handed lepton   superfield  ${\hat L}^i = ( {\hat N}_L,   {\hat
E}_L) $.
   In Table \ref{eqmot}
 we show these equations of motion, leaving off
the supersymmetry and gauge parts of the
variation.  Now it is easy
to see that these variations do remove parts of the cohomology
space that are proportional to the equations of motion.
 For example we have
\be {\cal A}' = \int d^4 x d^2 {\ov \q} \F^{i \a} c_{\a} {\hat {\ov J}}_i
{\hat {\ov E}}_R  =
\d  \{ \int d^4 x d^4 \q {\hat \F}^{i \a} \q_{\a} {\hat L}_i
 -  \int d^4 x d^2 {\ov \q} {\hat \F}^{i \a} c_{\a} {\hat {\tilde  {\ov
L}}}_i
\}
\ee
Without the use of the equation of motion, ${\cal A}'$  would
appear to be a supersymmetry anomaly of the theory.   It is clear
that one must calculate any real supersymmetry anomalies by
removing all such  spurious anomalies \ci{tamu46}.   Of course we
must translate the Higgs fields by their  vacuum expectation
values when using these equations of motion:  ${\hat J}^i \ra m  h
\d^i_1 + {\hat J}^i  ;
 \;\;  {\hat K}^i \ra m  h
\d^i_2 + {\hat K}^i $.
 The equations of
motion of the  antichiral source $\F'_{\a}$
should not be used, because it is not a
fundamental field and it does not propagate.

\begin{table}
$
\begin{array}{cc}
{\rm Variation} = & {\rm Equation \; of \;  Motion} \\
  \d {\hat {\tilde J}}_i = &   {\ov {\cal D}}^2 {\hat {\ov J}}_i  +
g_1  L_i E_R + g_2 K_i M - g_5 Q_i^c D_{R c} \\
  \d {\hat {\tilde L}}_i = &   {\ov {\cal D}}^2 {\hat {\ov L}}_i  -
  g_1  {\hat J}_i {\hat E}_R  \\
 \d {\hat  {\tilde E_R}} = &   {\ov {\cal D}}^2 {\hat {\ov E_R}}   +
 g_1  {\hat J}^i {\hat L}_i  \\
\d {\hat  {\tilde K}}_i = &   {\ov {\cal D}}^2 {\hat {\ov K}}_i  -
g_2  {\hat J}_i {\hat M} - g_4 {\hat Q}_i^{c} {\hat U}_{R c}\\
\d {\hat  {\tilde M}}_i = &   {\ov {\cal D}}^2 {\hat {\ov M}}   +
g_2  {\hat J}^i {\hat K}_i  + g_3 m^2\\
\d {\hat  {\tilde Q}}_{i c} = &   {\ov {\cal D}}^2 {\hat {\ov Q}}_{ic}   +
  g_4  {\hat K}_i {\hat U}_{R c} + g_5 {\hat J}_i
{\hat D}_{R c} \\
\d {\hat  {\tilde U_R}}^{  c} = &   {\ov {\cal D}}^2 {\hat {\ov U}}_R^c   +
g_4  {\hat Q}^{ic} {\hat K}_i  \\
\d {\hat  {\tilde D_R}}^{  c} = &   {\ov {\cal D}}^2 {\hat {\ov D}}_R^c   +
 g_5  {\hat Q}^{ic} {\hat J}_i
\end{array}$
\caption{Equations of Motion of Chiral Superfields}
\la{eqmot}
\end{table}

\section{Composite Superoperators and their Anomalies}

 A selection of candidate superanomalous operators
and their anomalies for the standard model can be
found in Table
\ref{examples} .  The labels
L,M,B,V,H stand for leptons, mesons, baryons, vector bosons and Higgs
respectively.  One of these particles can be found in each set.
Sometimes it will be in the operator and sometimes in the anomaly.

\begin{table}
$\begin{array}{cccccccc}
   {\rm Label } & {\rm Operator}
 & {\rm Anomaly}
 &  {\rm Dim} & {\cal L} & {\cal B}
 & {\cal Y} & {\cal I} \\
  L1    & {\cal D}^2   [ {\hat J}^j  {\cal D}_{ \a}     {\hat E}_R ]  & m^3
{\hat {\ov L}}^j  c_{\a}   &    \fr{7}{2}& -1 &
 0 & 1& \fr{1}{2} \\
 L2 & {\cal D}^2   [ {\hat J}^i   {\hat W}^j_{i \a}    {\hat E}_R ]   &
m^4    {\hat {\ov L}}^j  c_{\a}   &
\fr{9}{2}&  -1 &
 0 & 1 & \fr{1}{2} \\ L 3 & {\cal D}^2   [  {\hat J}^i {\hat W}^j_{i
\a}
    {  L}_j ]
 & m^4         {\hat {\ov E }}_R   c_{\a}    &    \fr{9}{2}&  1 &
 0 & -2 & 0 \\
 M1 & {\cal D}^2  [ {\hat U}_{R a}    {\cal D}_{\a}  {\hat Q}_{  i   }^a  ]
 & m^2   {\ov Q}_{   i  c}
    {\ov D}_{R  }^c  c_{\a} &
\fr{7}{2} &  0 &  0 &  -1& \fr{1}{2} \\
 B1 & {\cal D}^2  [ {\hat Q}^{i  a} {\hat Q}_{ i }^{ b} {\cal D}_{\a}   {\hat
Q}_{ j }^{c}\e_{abc} ]     &   m    {\hat {\ov  U}}_{R }^{  a }   {\hat {\ov
D}}_{R }^{ b}  {\hat {\ov  C}}_{R}^{   c}  {\hat {\ov  K}}_j
\e_{abc}  c_{\a}  & \fr{9}{2}&  0 & 1& 1 & \fr{1}{2}  \\ B2 &  {\cal D}^2  [
{\hat Q}^{i  a} {\hat Q}_{  i }^{ b}  {\hat K}^{j c} {\cal D}_{\a}   {\hat Q}_{
j }^{c}\e_{abc}] &  m^2    {\hat {\ov  U}}_{R }^{  a }   {\hat {\ov  D}}_{R
}^{ b}  {\hat {\ov  C}}_{R}^{   c}
\e^{abc}  c_{\a}  &   \fr{9}{2} & 0& 1&2& 0\\ B3 &  {\ov {\cal D}}^2  [
{\hat {Q}}^{i a}  {\hat {\ov D}}_{R}^{ b} {\ov {\cal D}}_{\dot
\a}   {\hat {\ov U}}_{R}^{c}\e_{abc} ]  & m^2    {\hat {Q}}^{ i a }   {\hat  Q
}^{    b}_j  {\hat {Q'}}^{ j  c}
\e_{abc}   {\ov c}_{\dot \a} &
\fr{9}{2}&  0& 1&   1& \fr{1}{2}
\\ B4 & {\ov {\cal D}}^2  [  {\hat {\ov U}}_{R }^{a} {\hat {\ov D}}_{R}^{ b}
{\ov {\cal D}}_{\dot \a}   {\hat {\ov U}}_{R}^{c}\e_{abc} ]  & m     {\hat
{Q}}^{ i a }   {\hat  Q }^{    b}_j  {\hat {Q'}}^{ j  c}
 {\hat {K}}_{ j  }
\e_{abc}   {\ov c}_{\dot \a} & \fr{9}{2} &  0&  1&  2 & 0 \\ V1 & {\cal
D}^2 [  {\hat W}^i_{j \a} {\hat J}_{i} ] & m^2 {\hat {\ov L}}_{L j} {\hat
{\ov E}}_{R} c_{\a} & \fr{7}{2} &  0&  0 & -1 &
\fr{1}{2} \\ H1 & {\cal D}^2 [  {\hat W}^i_{j \a} {\hat K}_{i} ]  & m^3
{\hat {\ov J}}_{  j} c_{\a} & \fr{7}{2} &  0&  0 &   1 &
\fr{1}{2} \\
\end{array}$
\caption{Some candidates for superanomalies  in
the SSM}
\la{examples}\end{table}

A comparison of Tables \ref{eqmot} and \ref{examples} reveals
that each of the above anomalies does represent a class that does
not vanish by the equations of motion.
It is also clear that to find the coefficient of
the anomaly, one needs to evaluate a large set of
diagrams and compare coefficients to see if they
can all be eliminated by the equations in Table
\ref{eqmot} or not.  So it looks much harder to
calculate these coefficients than it is for the
gauge anomalies, where mixing is not much of a
problem at all.  For example, one can add the
term $\int d^4 x
\e^{\m\n\l\s} V_{\m} A_{\n} \pa_{\l} V_{\s}$ to
the action for the VVA triangle anomaly to
conserve one type of current at the expense of
another, but this is not a major mixing problem
like we get for the superanomalies.

 The leptons occur in the anomaly in operators L1,L2 and L3.
The most important thing to notice is that the electron superfields
$E_L, E_R$ and  the neutrino superfield
$N_L$ appear linearly in these expressions
(see equation (\ref{lepanom1}) above).  This implies
that the elementary lepton fields are  being
mixed with the composite
field $\F_{\a}$ in a  non-supersymmetric way,
which will result in some mass splitting of both
multiplets (see equation (\ref{anomaction}) above), assuming  that
the coefficients of the anomalies are not zero.

 In  the quark model the $\Pi^-$ would naturally
 be produced by
such  operators as:
\be
\Pi^-   = {\ov u}_{L a  \dot   \a}
    {\ov d}_{R }^{  a \dot \a}
 \la{meson1}
\ee
The corresponding
anomaly would be M1.    The expansion of
${\ov Q}_{   i  c}
    {\ov D}_{R  }^c$
superfield gives the pion operator  above ( plus other
isospins and terms for supersymmetry) as its ${\ov F}$ term.

Let us   now write down some typical interpolating operators that
could  be used to create a proton:
\be  { P}_{L  \a   } = u_{L   \b}^{a} d_{L }^{ b \b}  u_{L    \a}^{c}
\e_{abc} \la{proton1} \ee
\be { P}_{M   \a} =  {\ov u}_{R   \dot
\b}^a {\ov  d}_{R  }^{b \dot \b} {  u}_{      \a}^c
 \e_{abc} \la{proton1a}\ee
\be { P}_{R  \dot \a} =  {\ov u}_{R   \dot \b}^a {\ov  d}_{R  }^{b \dot \b}
{\ov u}_{R   \dot \a}^c
 \e_{abc} \la{proton2}\ee
\be  { P}_{M  \dot \a   } = u_{L   \b}^{a} d_{L }^{ b \b}  {\ov u}_{R
\dot
\a}^{c}
\e_{abc} \la{proton2a} \ee

The supersymmetric generalization of the  familiar composite
operators (\ref{proton1})  are  to be found in
B1 and B2.   For      (\ref{proton2}) we have  B4.          The  third $Q'$
in the anomalies  must involve new flavours   so that the
expression does not vanish.     $C^a_R$ is the  right-handed charm
superfield, also needed so the anomaly does not vanish.
Evidently there is a lot of stucture here that involves the flavour
symmetry.

Now let us consider the weakly interacting vector
boson supermultiplets.  Here the  relevant observable states
 occur both in
the operator (V1)  and in the anomaly (H1).
 Recall that  in the SSM, the  Higgs
chiral superfields  combine
to form part of the supermultiplet that contains the
vector bosons.    Evidently, there is a lot of analysis to be done in
this sector too.  Also, the
cohomology of the vector superfields is not
yet known  completely, so there could be even more anomalies than
the ones that are discussed here.   But it does not appear necessary
to find more anomalies to have a good chance of breaking
supersymmetry in all the observed particles.

It is worthwhile to note that if there are superanomalies in  the
lowest dimension   operator with   given quantum numbers, this
will spread to all the higher dimensional   operators with those
quantum numbers.   The reason is that higher dimensional
operators   will necessarily mix with the
lower dimension operators with the same quantum numbers, with
coefficients which involve the appropriate power of mass.     So
supersymmetry breaking could not be rescued by using a different
interpolating operator.

Finally we note that the true    neutral
massive Higgs supermultiplet ${\hat H} = \fr{1}{\sqrt{2}} [{\hat J}^1
- {\hat K}_1]$   could also  appear in an
anomaly in a term like H1.  However the coefficient here may be
zero, for reasons discussed in section (\ref{index}) below.

\section{Higher Spins}

One possible   expression for the spin $\fr{3}{2}$ operator
corresponding to the $\D^{++}$ member of the baryon decuplet is \be
\D^{++}_{\a \b
\g} = u_{L
\a }^{a} u_{L }^{ b \b}  u_{L    \g}^{c} \e_{abc}
\la{delta}\ee
Is there any    chance of splitting the mass of this particle
from those of its superpartners in analogy with the above?

Evidently, this operator  occurs   in  the
following spin
$\fr{3}{2}$ superfield:
\be
\F^{(ijk)}_{(\a\b\g)} =
{\cal D}_{(\a}   {\hat Q}^{ i  a} {\cal D}_{\b} {\hat Q}^{j b} {\cal
D}_{\g )}   {\hat Q}^{k c}\e_{abc} ]
\ee
where the round brackets in $(ijk)$ and $(\a\b\g)$ denote
symmetrization.

Is there any anomaly available for this case? It appears quite
likely.    In
\ci{dmr}, it was shown that there are polynomials in the BRS
cohomology space for   spinor superfields of all
spins. The anomaly for spin
$\fr{3}{2}$ would have spin 1.  Reference \ci{dmr} needs to be
extended to the case where gauge fields are included to ensure
that this works out properly.

The  $\r^+$ meson corresponds to operators such as: \be
\r^+_{\a \dot \b}   = u_{L   \a}^{c}  {\ov d}_{L c \dot \b }
 \la{meson3} \ee  Particles such as these  will presumably appear
in the spin $1$ anomalies  of spin $\fr{3}{2} $   superfields
and perhaps also in M1 of Table {\ref{examples}.

Therefore it seems quite possible that there are superanomalies
available for all the myriad of hadronic resonances that have
been observed.  The question of course is whether nature makes
use of them, and that is most easily determined for the low spin
cases discussed above.

\section{The Witten Index and Superanomalies}
\la{index}
The lepton supermultiplets, the $0^-$ meson
supermultiplets and the Higgs supermultiplets all
occur as possible superanomaly terms for various
composite antichiral spinor superfields, chosen to
have the right quantum numbers. On the other
hand,  the $\fr{1}{2}^+$ baryons are naturally
written as composite antichiral spinor superfields
that can develop superanomalies with the appropriate
quantum numbers.   The  vector boson
supermultiplets  also can get mixed
with other operators through superanomalies.

Are  there any guidelines that can aid one to find a way through
all the complexity of the necessary calculations?  If there are
superanomalies, where  do  they come from?      Fortunately, there
is a very plausible and simple conjecture that naturally arises here,
although there is still much mystery connected with the details.

The well-known gauge and
gravitational anomalies  of quantum field theory
are associated with   index theorems for the
Dirac operator \ci{ginsalv}.   The anomaly measures the
appropriate index,  which is equal to  the
number of left handed fermion zero modes minus
the number of right handed fermion zero modes in
the relevant background gauge or gravitational
field.

If the  superanomalies discussed above do exist with non-zero
coefficients, it would be natural to expect that
they also measure an index. The obvious choice for
such an index is the Witten index \ci{witten}, which measures
the number of bosonic zero modes minus the number
of fermionic zero modes in these supersymmetric theories. For
present  purposes, the Witten  index  would  have to be defined for
each relevant zero mass supermultiplet separately, since different
superanomalies would evidently measure different Witten
indices.

Thus     the Witten
indices of the neutrino and   photon superfields
must be associated with  the weak-EM type superanomalies L1,
L2,  L3, V1 and H1 of  Table \ref{examples}. The hadronic
superanomalies M1 and B1-B4 would be associated with  the Witten
indices of the gluon  and  photon superfields.

It is well known and also clear from the Feynman
diagrams that the supersymmetry BRS identities
work by cancellation of the fermionic and bosonic
modes.  From the results of \ci{witten}, we know that these can be
unmatched only for zero mass fields at zero momentum and  in fact
frequently are unmatched in these very circumstances.
In the context
of looking for non-perturbative spontaneous breaking
of supersymmetry, this was a disappointing
result,  because a theory with a non-zero Witten
index cannot break supersymmetry spontaneously \ci{witten},
even beyond perturbation theory-- it necessarily must possess
some zero-energy state.  However, these very  theories that cannot
break supersymmetry spontaneously are the theories
which are most likely to  possess superanomalies,
precisely because  they have non-zero Witten
indices.

This leads to a  puzzle.   In  reference \ci{coleman} it is
demonstrated that   `the invariance of the vacuum
is the invariance of the world'.   This means that if the vacuum is
supersymmetric, and it must be if the Witten index is non-zero,
then all the states must be in supermultiplets,
even non-perturbatively.  If this theorem applies to the
 present situation, it means that  superanomalies can not possibly
break the supersymmetry in a theory with non-zero Witten
index, even if one goes beyond perturbation theory.

However, one of the  explicit assumptions in the theorem of
reference  \ci{coleman} is that there are  no
zero mass particles, and of course these are  precisely the origin of
the superanomalies.  So theorem \ci{coleman} does not apply to
our situation! It appears to be  possible to have a supersymmetric
vacuum without having  the states in  supermultiplets precisely in
the case when the total Witten index is non-zero.  There is a bonus
involved in having the vacuum supersymmetric of course--the
cosmological constant is exactly zero even after supersymmetry
breaking, and this statement survives even beyond perturbation
theory!

But the comparison with the gauge anomalies raises
another issue.  The zero modes are related to the gauge
and gravitational anomalies in situations where the topology of
the gauge fields is non-trivial, e.g. in the field of an instanton.
So perhaps we should be looking at topologically non-trivial field
theories for supersymmetry anomalies--particularly in theories
where the topology of the Higgs sector chiral
superfields   is non-trivial, as in the `t Hooft-Polyakov
magnetic monopole solution for example.  In fact, there is
evidence that supersymmetry has problems with zero modes in
this very situation \ci{casher}.

 All of the above discussion applies only to the rigidly
supersymmetric theory of course. Obviously there is no  unitarity
problem created by the superanomalies in this case, because it is
only a rigid symmetry that is violated.

What happens when we couple the theory to a supergravity
theory derived from a superstring?   On the positive side, this
will generate a zero  cosmological constant, so long as
supersymmetry is not spontaneously or explicitly broken.   Does
the above mechanism for supersymmetry breaking survive with a
zero cosmological constant?  In fact it does appear to have a
chance of doing so, because  we do not want to couple
  operators like those in Table \ref{examples}   to
the supergravity theory anyway-- it would be wrong to couple,
say, both a fundamental  proton   operator and the quarks to
gravity in the same theory.    The basic idea would be that the
scale of gravity is so huge compared to the scale of the leptons
and hadrons that we should be dealing with the rigid theory when
considering the observable particles anyway.  It may therefore be
comforting, rather than disappointing, that it is very difficult to
couple  antichiral spinor superfields $\F_{\a}$  to   supergravity
theory.   So the supergravity would still be unitary and
supersymmetric--this appears to imply that the gravitino should
be massless.

It is remarkable that  for nearly all of the examples in this
standard model, there are in fact zero mass
fields which do contribute to the Feynman diagrams
which can take one from the operator  to the
superanomaly.   This is the main reason why it seems best to look
for these anomalies in the context of   the standard model.
Simpler examples that look as promising are hard to find
and not so interesting.    But note that the Higgs  particle
  does not couple to any massless field in the rigidly
supersymmetric theory.  So the Higgs supermultiplet   may not
be involved in any superanomaly.  Assuming that
the scheme works as outlined above, this has
the   surprising consequence that  the Higgs
particles should remain in a supermultiplet.

\begin{center}
Acknowledgments
\end{center}

It is a pleasure to recall many useful remarks made to me   in the
course of this work by R. Arnowitt, M. Duff, S. Ferrara, R. Khuri, U.
Lindstrom, R. Minasian,   H. Pois, S. Polyakov,  C. Pope, J. Rahmfeld,
P. Ramond, M. Rocek,   E. Sezgin, K. Stelle, C. Thorn,  and E. Witten.

\end{document}